\documentclass[prd,twocolumn,amsmath,amssymb,aps]{revtex4-2}
\makeatletter
\newcommand*{\rom}[1]{\expandafter\@slowromancap\romannumeral #1@}
\makeatother
\usepackage{graphicx}
\usepackage{dcolumn}
\usepackage{bm}
\usepackage[noend]{algpseudocode}
\usepackage[ruled,vlined]{algorithm2e}
\usepackage{amsmath,xparse,mleftright}
\usepackage{subfigure}
\usepackage{braket,color}
\usepackage{dsfont}
\usepackage[table]{xcolor}
\usepackage{tensor}
\usepackage{float}

\begin{document}

\title{A correspondence between Maxwell--Einstein theory and superfluidity}%

\author{Emil Génetay Johansen}
\affiliation{Optical Sciences Centre, Swinburne University of Technology, Melbourne 3122, Australia}
\email[]{e.genetay\_johansen@unsw.edu.au}

\begin{abstract}
A planar superfluid is considered and interpreted in terms of electromagnetism and gravity. It has previously been suggested that the superfluid flow can be regarded as analogous to an electromagnetic field and that a non-vanishing density gradient give rise to a gravity-like force. The present work seeks to reconcile these hitherto distinct pictures into a unified exposition of vortex electrodynamics in a curved analogue space-time. By constructing a theory in which the planar Maxwell's equations are coupled to a $(d+1)$-dimensional conformally curved space-time, we expose a correspondence between the resulting equations of motion of the embedded fields and the dynamics of a quantum fluid. Finally, an effective vortex metric whose connection components are torsional is studied and its effect on the superfluid Maxwell's equations is elucidated upon.

\end{abstract}

\maketitle

\section{Introduction}

Einstein's theory of gravity \cite{albert1916foundation} and Maxwell's theory of electromagnetic fields \cite{1865RSPT..155..459C} may be regarded as the cornerstones of classical physics. While general relativity establishes a relationship between energy and space-time geometry, electromagnetic field theory furnishes a description of how charged particles interact, and gives rise to the notion of light. In pioneering works put forth by Schwinger \cite{PhysRev.73.416,PhysRev.74.1439}, Tomonaga \cite{1946PThPh...1...27T} and Feynman \cite{PhysRev.76.769,PhysRev.76.749,PhysRev.80.440}, Maxwell's theory was successfully quantized and unified with special relativity, which culminated in a fully relativistic theory of quantum electrodynamics (QED). This crowning achievement marked the beginning of the twentieth century revolution in elementary particle physics. The ideas upon which QED was founded were further developed by considering more general unitary gauge groups $\rm{SU}(N)$. Such theories, collectively known as Yang--Mills theories \cite{PhysRev.96.191,RevModPhys.52.661,Catren2008GeometricFO}, are at the heart of the Standard Model of elementary particles and our contemporary understanding of fundamental physics.

Gravity is significantly weaker than the forces emerging from these theories though. While the ratio between the strong nuclear force and electromagnetism is about $10^{-2}$, and electromagnetism and the weak nuclear force about $10^{-4}$, the gap between the weak nuclear force and gravity is approximately $10^{-33}$. Explaining this vast discrepancy constitutes one of the central problems in foundational physics and is known as the hierarchy problem. In the vast majority of instances, the impact of gravity on electromagnetism, and conversely, the effects of electromagnetism on gravity, is negligible; gravity is generally too weak to significantly perturb the electromagnetic fields, and the energy-momentum carried by the electromagnetic fields is most often too low to cause a measurable curvature of space-time. The main exception to this approximation is of course in the context of black holes. It is well established that the strong gravitational pull by a black hole is causing light passing by in its vicinity to deflect \cite{1979Natur.279..381W,2008AHES...62....1S,1919Natur.104..372L}. In such a scenario, and in situations alike, electromagnetic field theory and gravity cannot be treated independently since the coupling between the two is non-negligible. This is generally also the case when gravity and electromagnetism is simulated in a superfluid. It was demonstrated by Popov \cite{popov1973quantum} that a superfluid can be described in terms of relativistic electrodynamics where the superfluid flow corresponds to the the electric field and the condensate density to the magnetic field. In addition, Unruh showed \cite{PhysRevLett.46.1351} that excitations in a superfluid experience the background system as an effective space-time where a non-vanishing density gradient accounts for the curvature \cite{PhysRevLett.46.1351,1998CQGra..15.1767V,2000PhRvL..85.4643G,2002PhRvD..66d4019S,2005GReGr..37.1549W,2007PhRvA..76c3616J,2013JPCM...25N4211H,2014AcPPB..45..905C,2004gr.qc.....1025F,1998JETPL..67..698V}. Any density landscape can in principle be imprinted with the technologies available today \cite{gauthier2016direct}, so the effective gravity in a superfluid may well be strong enough to affect the properties of the superfluid electromagnetic fields, just as when light is propagating in the presence of a black hole. This calls for an extension of the vortex electrodynamics picture to a complete formulation accounting for this interaction.\\
Over the years, several strategies for unifying electromagnetism and gravity have been proposed including the so-called teleparallel formulation \cite{RG,2014shst.book..197P} developed by Einstein, Kaluza--Klein theory \cite{2018IJMPD..2770001K,1926ZPhy...37..895K,1926Natur.118..516K,Pasini1988ACI} and its precursor, Nordström's theory \cite{Nordstroem2007,nordstrom1913trage,nordstrom1913theory}, and in addition Maxwell--Einstein theory \cite{2005CQGra..22..393T,Papapetrou1974}.
Teleparallelism relies on the notion of a superpotential and is constructed in a non-coordinate basis via the introduction of tetrad fields \cite{deFelice:1990hu}, which makes possible a description of gravity that is quadratic in field strength, akin to Yang--Mills. Kaluza--Klein theory and Nordström's theory share the common trait that gravity and electromagnetism are treated as two parts of the same object and in Maxwell--Einstein theory the space-time curvature is viewed as a background gauge-like field to which the field strength tensor is coupled. It may thus be natural to pose the question of whether the vortex electrodynamics description can be naturally reconciled with analogue gravity into a single theory. The present work is devoted to addressing this particular question. Finally, for the sake of completeness, we also consider a vortex metric whose connection components have non-vanishing torsion.

\section{Vortex electrodynamics}

The wavefunction $\Psi(\text{r},t)$ describing the collective behaviour of a weakly interacting superfluid is a solution to the Gross--Pitaevskii equation (GPE) \cite{dalfovo1999theory}
\begin{equation}
\label{GPE}
   i\hbar \frac{\partial \Psi(r,t)}{\partial t} = \left(-\frac{\hbar^2}{2m}\nabla^2 + g |\Psi(r,t)|^2 - \mu\right)\Psi(r,t),
\end{equation}
where $m$ is the mass of the condensate particle, $g$ is the coupling constant and $\mu$ is the chemical potential. By performing a Madelung transformation $\Psi(\text{r},t) = |\Psi(\textbf{r},t)| e^{i S(\textbf{r},t)}$, the GPE can be split into an Euler-like equation
\begin{equation}
\label{euler}
     -\hbar \frac{\partial S}{\partial t} = \frac{m}{2} v_{\rm{sf}}^2 + g|\Psi(\textbf{r},t)|^2 - \mu
\end{equation}
and a continuity equation 
\begin{equation}
\label{cont}
    \frac{\partial n(\textbf{r},t)}{\partial t} = \nabla \cdot \left(n(\textbf{r},t) \boldsymbol{v}_{\rm{sf}}\right),
\end{equation}
where $n(\textbf{r},t) = |\Psi(\textbf{r},t)|^2$ and $\boldsymbol{v}_{\rm{sf}}(\textbf{r},t) = \frac{\hbar}{m}\nabla S(\textbf{r},t)$. In Ref.~\cite{2020PhRvA.101f3616S}, it was shown that the following definitions for the superfluid electric and magnetic field
\noindent\begin{minipage}[h]{.5\linewidth}
\begin{equation}\nonumber
  \textbf{E}_{\rm{sf}} = m n_0 \boldsymbol{v}_{\rm{sf}} \times \hat{\textbf{e}}_z
\end{equation}
\end{minipage}%
\begin{minipage}[h]{.5\linewidth}
\begin{equation}
  \textbf{B}_{\textrm{sf}} = \frac{\hbar m}{g} \partial_t S \hat{\textbf{e}}_z
\end{equation}
\end{minipage}
with vacuum constants

\noindent\begin{minipage}{.5\linewidth}
\begin{equation}\nonumber
  \epsilon_{\rm{sf}} = \frac{1}{m n_0}
\end{equation}
\end{minipage}%
\begin{minipage}{.5\linewidth}
\begin{equation}
 \mu_{\rm{sf}} = \frac{m^2}{g}
\end{equation}
\end{minipage}
such that the speed of sound satisfies the square root law 
\begin{equation}
    c_{\rm{sf}} = \frac{1}{\sqrt{\epsilon_{\rm{sf}} \mu_{\rm{sf}}}},
\end{equation}
solve Maxwell's equations under the assumption that the density $n(\textbf{r}) = n_0$ is constant. There are, however, phonons in the system which in this formulation play the role of photons, which are linear perturbations of this otherwise flat background. This particular configuration is just one of an infinite set of possible solutions though. In general, it holds true that any configuration
\noindent\begin{minipage}{.5\linewidth}
\begin{equation}\label{fields}\nonumber
  \textbf{E}_{\rm{sf}} = m n_0^{x+1} \boldsymbol{v}_{\rm{sf}} \times \hat{\textbf{e}}_z
\end{equation}
\end{minipage}%
\begin{minipage}{.5\linewidth}
\begin{equation}
  \textbf{B}_{\textrm{sf}} = \frac{\hbar m n_0^x}{g} \partial_t S \hat{\textbf{e}}_z
\end{equation}
\end{minipage}
with vacuum constants

\noindent\begin{minipage}{.5\linewidth}
\begin{equation}\label{constants}\nonumber
  \epsilon_{\rm{sf}} = \frac{1}{m n_0^{x+1}}
\end{equation}
\end{minipage}%
\begin{minipage}{.5\linewidth}
\begin{equation}
 \mu_{\rm{sf}} = \frac{m^2 n_0^{x}}{g}
\end{equation}
\end{minipage}
forms a solution for any arbitrary constant $x \in \mathds{R}$. We shall later see that this constant has to be an integer in a Maxwell--Einstein formulation as it relates to the dimension of the pertinent space-time. Now, let us generalize the above results by allowing for a variable density $n(\textbf{r})$ while keeping it constant in time. 
\subsection{Generalized Gauss-like law}

Gauss's law in the non-covariant formulation is denoted as
\begin{equation}
    \nabla \cdot \textbf{E}_{\rm{sf}} = \frac{\rho_v}{\epsilon_{\rm{sf}}},
\end{equation}
where $\rho_v$ is the vortex density. Inserting the electric field in Eq.~\eqref{fields} and expanding the divergence on the left hand side we obtain after some rearrangements
\begin{equation}
    n^{x+1} \left(\partial_{r} + (x+1) \frac{\partial_r n}{n}\right)v_{\rm{sf}} = \frac{\rho_v}{\epsilon_{\rm{sf}}} + \lambda(\textbf{r}),
\end{equation}
where $\lambda(\textbf{r})$ is an additional source term accounting for the interaction between electromagnetism and gravity. The physical interpretation of this term will be clarified in the next section. Now, if we divide both sides by $n^{x+1}$ and write the superfluid flow as $v_{\rm{sf}} = n^{-(x+1)} (n^{(x+1)} v_{\rm{sf}}) = {n^{-(x+1)}} \textbf{E}_{\rm{sf}}$, we arrive at
\begin{equation}\label{gengauss}
    \left(\partial_{r} + (x+1) \frac{\partial_r n}{n}\right)n^{-(x+1)}\textbf{E}_{\rm{sf}} = n^{-(x+1)}\frac{\rho_v}{\epsilon_{\rm{sf}}} + \Tilde{\lambda}(\textbf{r}),
\end{equation}
where the factor $n^{-(x+1)}$ was absorbed into $\Tilde{\lambda}(\textbf{r})$. The resulting equation has the same form as an electric field (scaled by ${n^{-(x+1)}}$) minimally coupled to a gauge field with coupling $x+1$. This field is nothing but the velocity $\boldsymbol{v}_{\nabla |\Psi|}$ due to the density gradient
\begin{equation}
\label{v_d}
    \boldsymbol{v}_{\nabla |\Psi|}(\textbf{r}) = \frac{\hbar}{m} \frac{\nabla n(\textbf{r})}{n(\textbf{r})}.
\end{equation}
Focusing on the first term $\partial_{r} {n^{-(x+1)}}\textbf{E}_{\rm{sf}}$ a straight forward calculation gives us the vortex density
\begin{equation}
    \partial_{r} v_{\rm{sf}} = \nabla \times \boldsymbol{v}_{\rm{sf}} \cdot \hat{\textbf{e}}_z = \rho_{v},
\end{equation}
since $n^{-(x+1)} \epsilon_{\rm{sf}}^{-1} = m$. The second term proportional to the density gradient arises due to the fact that the vortex is moving with a speed $v_{\nabla |\Psi|}$. Thus, since the conventional Gauss's law only holds in the rest frame of the vortex, we must perform a frame transformation into this frame, which results in a coupling to the gauge field $v_{\nabla |\Psi|}$. The source of this interaction can therefore be identified as
\begin{equation}
    \Tilde{\lambda}(\textbf{r}) = m(x+1) v_{\nabla |\Psi|}(\textbf{r}) v_{\rm{sf}}(\textbf{r})
\end{equation}

\subsection{Generalized Ampère-like law}

Ampère's law in its non-covariant form can be written
\begin{equation}
    \nabla \times \textbf{B}_{\rm{sf}} = \frac{\partial \textbf{E}_{\rm{sf}}}{\partial t} + \mu_{\rm{sf}} \textbf{j}_v,
\end{equation}
where $\textbf{j}_v = \boldsymbol{v}_v \rho_v$ denotes the vortex current, and $\boldsymbol{v}_v$ is the velocity of the source vortex. Going through the same steps as in the preceding subsection, we find
\begin{equation}\label{genamp}
    \left(\partial_{\theta} + x \frac{\partial_r n}{n}\right)n^{-x} \textbf{B}_{\rm{sf}} = n^{-x} \left(\epsilon_{\rm{sf}}\mu_{\rm{sf}}\frac{\partial \textbf{E}_{\rm{sf}}}{\partial t} + \mu_{\rm{sf}} \textbf{j} + \hat{\lambda}(\textbf{r})\right).
\end{equation}
Again, we start off by focusing on the first term on the left hand side, which gives us 
\begin{equation}
    \partial_{\theta} \left(\frac{\hbar m}{g}\partial_t \theta\right) = \frac{\hbar m}{g} \partial_t \left(\partial_{\theta} S\right) = \epsilon_{\rm{sf}}\mu_{\rm{sf}}\frac{\partial \textbf{E}_{\rm{sf}}}{\partial t}.
\end{equation}
The current $\textbf{j}_v$ can be obtained by simply performing a frame transformation with respect to $v_v$ \cite{2020PhRvA.101f3616S}, not taking into account $v_{|\nabla \Psi|}$. In the rest frame of the vortex with speed $v_v$, we thus get
\begin{equation}
    i \hbar \partial_t \Psi = \left(i \hbar \partial_t - \textbf{J}_v \cdot \textbf{A}\right) \Psi_v,
\end{equation}
where $\textbf{J}_v \cdot \textbf{A} = -m \textbf{j}_v \cdot \textbf{r} \times \hat{\textbf{e}}_{z}$. Now, taking the curl of this expression we get the definition of the vortex current
\begin{equation}
    \partial_{\theta}\left(m \textbf{j}_v \cdot \textbf{r}\right) \hat{\textbf{e}}_{\theta} = \nabla \left(m \textbf{j}_v \cdot \textbf{r}\right) \times \hat{\textbf{e}}_{z} = \mu_{\rm{sf}} \textbf{j}_v.
\end{equation}
We also need to perform an additional frame transformation with respect to $v_{\nabla |\Psi|}$. Doing this, the source accounting for the minimal coupling is found via identification as
\begin{equation}
    \hat{\lambda}(\textbf{r}) = \frac{\hbar m}{g} v_{\nabla |\Psi|} \left(\partial_t S + \frac{m}{\hbar} \textbf{j}_v \cdot \textbf{r} \times \hat{\textbf{e}}_z\right)\hat{\textbf{e}}_z.
\end{equation}
\subsection{Generalized Faraday-like law and Gauss-like \textbf{B} law} \label{FG}

The superfluid Faraday-like law is given by
\begin{equation}
    \nabla \times \textbf{E}_{\rm{sf}} = - \frac{\partial \textbf{B}_{\rm{sf}}}{\partial t}.
\end{equation}
Taking the two-dimensional curl of the $\textbf{E}_{\rm{sf}}$ field gives us
\begin{equation}
    \nabla \times \textbf{E}_{\rm{sf}} = -m n^{x+1} \partial_{\theta} v_{\rm{sf}} \hat{\textbf{e}}_{z},
\end{equation}
and as for the right hand side, using the Euler-like Eq.~\eqref{euler} and the continuity equation Eq.~\eqref{cont}, we get after applying the time derivative
\begin{equation}
   \frac{\partial \textbf{B}_{\rm{sf}}}{\partial t} = -m\left( xn^x\partial_{\theta} v_{\rm{sf}} +n^{x+1} \partial_{\theta} v_{\rm{sf}} \right) \hat{\textbf{e}}_{z}.
\end{equation}
Since we are only considering static density landscapes, this equation must be satisfied due to the continuity equation $\partial_t n = - m n \partial_{\theta} v_{\rm{sf}}$. Finally, let us look at the Gauss-like $\textbf{B}$ law. This equation is, just as in the flat case, trivially satisfied since the planar divergence of a vector pointing in a direction orthogonal to the plane must vanish, i.e.
\begin{equation}
    \nabla \cdot \textbf{B}_{\rm{sf}} = \nabla \cdot \frac{\hbar m n^x}{g} \partial_t S \hat{\textbf{e}}_{z}=0,
\end{equation}
A noteworthy remark regarding these two equations is that they seem to be insensitive to the density distribution. In fact, they are pure geometrical statements, which together, are commonly referred to as the Bianchi identity. In the language of differential forms \cite{2017grav.book.....M}, the source-full and source-less Maxwell's equations can, respectively, be formulated as 
\begin{align}
 d\textbf{F}=0 \\
 \star d \star \textbf{F} = \textbf{J},
\end{align}
where 
\begin{align}
    \textbf{F} = E_1 dt \wedge dx + E_2 dt \wedge dy + E_3 dt \wedge dz + \nonumber \\ B_1 dy \wedge dz + B_2 dz \wedge dx + B_3 dx \wedge dy 
\end{align}
is the field strength 2-form, $\star$ denotes the Hodge operator and $\textbf{J}$ is the source. Since the Hodge operation is carried out with respect to the underlying metric, the sourceful equation cannot be invariant under the change of metric. This is not true for the sourceless equation though, since the exterior derivative is coordinate free, thus implying a coordinate independence. In a superfluid, the spatial dependence of the density can be interpreted as an effective space-time curvature, meaning that the effective metric is determined by the density. This is the picture we shall adopt in the next section.

\section{A Maxwell--Einstein formulation}

The equations obtained in the preceding section appear reminiscent of a Maxwell theory minimally coupled to a gauge field $v_{\nabla |\Psi|}$. It is therefore conceivable that the same set of equations may be obtained via a Maxwell--Einstein formulation in which the electromagnetic field strength tensor is minimally coupled to analogue gravity. In this section we endeavour to demonstrate that this is the case. The basic building blocks we use in our construction are given by $\{g_{\mu \nu}(n), \Gamma^{\lambda}_{\mu \nu}(n), F_{\mu \nu}(n,\nabla,S), J^{\rm{ME}}_{\mu}(n,\nabla,S),g_i\}$, where $g_{\mu \nu}$ is the effective metric, $\Gamma^{\lambda}_{\mu \nu}(n)$ is the space of Christoffel symbols corresponding to $g_{\mu \nu}$, $F_{\mu \nu}$ is the superfluid field strength tensor, $J^{\rm{ME}}_{\mu}$ is the Maxwell--Einstein current, $g_i$ are the couplings and $n=n(\textbf{r})$ is the density. The metric for a $(d+1)$-dimensional conformal analogue space-time is given by \cite{2005LRR.....8...12B}
\begin{equation}\label{metric}
    g_{\mu \nu}(r) = \Omega(r)^{\frac{2}{d-1}} \left( 
\begin{array}{c|c} 
  -(c_{\rm{sf}}^2-v_{\rm{sf}}^2) & -v_i \\ 
  \hline 
  -v_j & \delta_{ij} 
\end{array} 
\right),
\end{equation}
where $\Omega(r)$ is the conformal factor, which is proportional to the density $n(r)$. In what follows, we shall consider two special cases of this metric: i) the metric describing the situation in which the vortices are kept far away from one another so that off-diagonal elements vanish, but with arbitrary density profile, and ii) the metric corresponding to the analogue space-time in the neighborhood of a vortex residing in a system with an otherwise constant density, such that the torsion is non-vanishing. Moreover, the $(2+1)$-dimensional field strength tensor is given by
\begin{equation}\label{F}
    F_{\mu \nu} = \left( 
\begin{array}{ccc}
  0 & \frac{E_{r}}{c_{\rm{sf}}} & \frac{E_{\theta}}{c_{\rm{sf}}} \\ 
  -\frac{E_r}{c_{\rm{sf}}} & 0 & B_{z} \\
  -\frac{E_{\theta}}{c_{\rm{sf}}} & -B_z & 0
\end{array} 
\right).
\end{equation}
In order to couple Maxwell's equations to gravity, we adopt the covariant formulation and compute the covariant derivatives $\nabla_{\mu}$ taking into account the change of coordinates due to the non-trivial structure of the effective space-time. Maxwell's equations, in this formulation, may then be written as
\begin{equation}\label{maxwell}
    \nabla_{\mu} F^{\mu \nu} = \mu_v J^{\nu}_{\rm{ME}},
\end{equation}
where 
\begin{align}
    F^{\mu \nu} = g^{\mu \alpha}g^{\nu \beta} F_{\alpha \beta} \label{raiseF} \\
    J^{\nu}_{\rm{ME}} = g^{\nu \alpha} J^{\rm{ME}}_{\alpha} \label{raiseJ}
\end{align}
and $J^{\rm{ME}}_{\alpha}=(c_{\rm{sf}}\rho_{\rm{ME}},\textbf{j}_{\rm{ME}})+\delta_{\alpha}(\textbf{r})$ are the covariant components of the Maxwell--Einstein current and $\delta_{\alpha}$ are the components of the correction term which are to be identified. The covariant derivative applied to the rank-2 tensor $F^{\mu \nu}$ takes on the form
\begin{equation}
    \nabla_{\mu} = \partial_{\mu} + g\left(\Gamma^{\alpha}_{\lambda \mu} + \Gamma^{\nu}_{\lambda \mu}\right),
\end{equation}
where $g$ is the coupling between gravity and the field strength tensor. The connection coefficients, in absence of torsion, 
 are given by the Christoffel symbols
\begin{equation}\label{CS}
    \Gamma^{\lambda}_{\mu \nu} = g^{\lambda \alpha} \left(\partial_{\nu}g_{\alpha \mu} + \partial_{\mu}g_{\alpha \nu} - \partial_{\alpha} g_{\mu \nu}\right).
\end{equation}
Note that due to the interchange symmetry of the lower two indices of $\Gamma^{\lambda}_{\mu \nu}$, the field strength tensor is left invariant so that
\begin{align}\label{invariant}
    F_{\mu \nu} = \nabla_{\mu} A_{\nu} - \nabla_{\nu} A_{\mu} =\nonumber \\ \partial_{\mu} A_{\nu} - \partial_{\nu} A_{\mu} + \left(\Gamma^{\alpha}_{\mu \nu} - \Gamma^{\alpha}_{\nu \mu}\right)A_{\alpha} \\
    = \partial_{\mu} A_{\nu} - \partial_{\nu} A_{\mu}, \nonumber
\end{align}
where $A_{\mu}$ are the components of the 3-vector potential. However, as we shall see in the next section, this is not the case when torsion is present, which thus necessitates the introduction of a more general connection, which is defined by Eq.~\eqref{torsioncon}.

\subsection{Non-zero curvature and vanishing torsion}

We shall now consider an arbitrarily conformally curved space-time in which the parallel transport is determined by the symmetric connection in Eq.~\eqref{CS}.

\subsubsection{Gauss--Einstein law}
Generalized Gauss's law can be obtained by setting $\nu=t$ in Eq.~\eqref{maxwell} which leads to the equation
\begin{align}
    \nabla_{\mu} F^{\mu t} = \left(\partial_r + \Gamma^{r}_{r r} + \Gamma^{t}_{t r} + \Gamma^{\theta}_{r \theta}\right)F^{r t}\nonumber\\ + \left(\partial_{\theta} + \Gamma^{\theta}_{\theta \theta} + \Gamma^{t}_{t \theta} + \Gamma^{r}_{\theta r}\right)F^{\theta t} = \mu_0 J^t_{\rm{ME}}.
\end{align}
Computing the relevant Christoffel symbol components and using Eq.~\eqref{raiseF} to obtain the contravariant field strength components results in the equation
\begin{equation}
    \nabla_{\mu} F^{\mu t} = -\left(\partial_r + g_E \frac{\partial_r n}{n}\right)n^{-\frac{d+1}{d-1}}F^{t r},
\end{equation}
where $g^{tt}g^{rr} = -\frac{d+1}{d-1}$. Moreover, we may compute the contravariant components of the current by means of Eq.~\eqref{raiseJ}, which yields
\begin{equation}
  J^t_{\rm{ME}} = g^{tt} \left(J^{\rm{ME}}_t + \delta_t\right) = n^{-\frac{d}{d-1}}(c_{\rm{sf}}\rho^{\rm{ME}} + \delta_t).
\end{equation}
We find that the resulting equation is in exact agreement with the generalized Gauss's law in Eq.~\eqref{gengauss}, given that we define the Maxwell--Einstein source and identify the correction as 
\begin{align}
    \rho_{\rm{ME}} = n^{-\frac{2-d}{d-1}} \rho_v \\  
    \delta_t = n^{-\frac{2-d}{d-1}}\Tilde{\lambda},
\end{align}
and enforce the following relationship between the spatial dimension $d$, the parameter $x$ and the coupling $g_E$
\begin{align}\label{xd}
    x=\frac{2}{d-1} \\
    g_E = x+1 = \frac{d+1}{d-1}.
\end{align}
\subsubsection{Ampère--Einstein law}
Next we consider the Ampère-Einstein law. Here we let $\nu = a$, where $a$ is a spatial index. Going through the same steps as for $\nu = t$, the left hand side of Eq.~\eqref{maxwell} becomes
\begin{equation}
    \nabla_{\mu} F^{\mu r} = -n^{-\frac{2}{d-1}} \partial_t F^{t r} + \left(\partial_{\theta} + g_M \frac{\partial_r n}{n}\right) n^{-\frac{2}{d-1}}F^{\theta r},
\end{equation}
since the $a=\theta$ equation vanishes. Again we recover the left hand side of the generalized Ampère's law in Eq.~\eqref{genamp} given the same relationship between $x$ and $d$ as in Eq.~\eqref{xd}, but now with the coupling $g_M=x$. Moreover, the Maxwell--Einstein current and the correction can be identified as
\begin{align}
    J^{r}_{\rm{ME}} = n^{-\frac{1}{d-1}} j_v \\
    \delta_r  = n^{\frac{1}{d-1}} \hat{\lambda}_r
\end{align}
for which the generalized Ampère's law derived from hydrodynamics is obtained.

\subsubsection{Einstein--Faraday law and Einstein--Gauss \textbf{B} law}
The Faraday and Gauss-like $\textbf{B}$ laws may, as already discussed, be regarded as geometric identities that cannot be obtained from Noether's theorem since there is no current to couple the fields to in the action. In the hydrodynamic picture, these were trivially invariant under the replacement $n \longrightarrow n(\textbf{r})$, which in a gravity picture entails that there is no response to the curvature of space. In the covariant formulation, these two equations may be summarized as
\begin{equation}
    \frac{1}{2} \partial_{\mu} \epsilon^{\mu \nu \gamma \delta}F_{\gamma \delta} = \partial_{\gamma} F^{\mu \nu} + \partial_{\mu} F^{\nu \gamma} + \partial_{\nu} F^{\gamma \mu} = 0.
\end{equation}
In order to show that this equation is insensitive to the underlying metric, we may first lower the indices of the field strength tensor by applying the metric. Doing this leads to an identical equation but for the covariant components instead of the contravariant ones, since the right hand side is zero. Replacing the derivatives with their covariant counterparts, and collecting the terms, thus gives us
\begin{align}\label{FGeq}
    \left(\partial_{\gamma} F_{\mu \nu} + \partial_{\mu} F_{\nu \gamma} + \partial_{\nu} F_{\gamma \mu}\right) + \left(\Gamma^{\alpha}_{\mu \gamma} - \Gamma^{\alpha}_{\gamma \mu}\right) F_{\alpha \nu}\\ \nonumber + \left(\Gamma^{\beta}_{\nu \gamma} - \Gamma^{\beta}_{\gamma \nu}\right) F_{\beta \mu} + \left(\Gamma^{\delta}_{\nu \mu} - \Gamma^{\delta}_{\mu \nu}\right) F_{\delta \gamma}=0.
\end{align}
Now, since the Christoffel symbols are symmetric under the interchange of the lower indices, all terms due to the curvature cancel out and we are left with the equation defined on a flat background, which defines the Bianchi identity and thereby must vanish too. Note, however, that this is only due to the fact that the Christoffel symbol is a symmetric connection, which means that it does not take into account the effects due to torsion. As a consequence of this, it is generally the case that the homogenous equations do not remain invariant under the change of metric when torsion is present. We shall return to this in the following section. We have now worked out the Maxwell--Einstein equations for the metric given by Eq.~\eqref{metric} and found that they are in exact agreement with the generalized Maxwell's equations derived from hydrodynamics, where $\textbf{E}_{\rm{sf}}$ and $\textbf{B}_{\rm{sf}}$ are defined as in Eq.~\eqref{fields} and the vacuum constants are given by Eq.~\eqref{constants}, provided the relationship between the parameter $x$ and the spatial dimension $d$ established in Eq.~\eqref{xd}, and couplings $g_E=x+1$ and $g_M=x$.

\subsection{Non-zero torsion and vanishing curvature}\label{torsion}

Let us now consider instead a system in which the density is kept constant but with a vortex residing in it. Thus, in the vicinity of the vortex, the metric defined in Eq.~\eqref{metric} is no longer diagonal, which consequently gives rise to a different Maxwell--Einstein theory that is coupled to the torsion of the underlying connection, whose equations of motion are derived in this section. However, before we go on and work these equations out, we shall outline some pertinent remarks regarding torsion and its relationship to vorticity.

\subsubsection{On torsion --- general remarks}
The vanishing of torsion and Lorentz invariance of $\eta_{\mu \nu}$ uniquely determine the connection to be of a Levi--Civita kind, since the manifold must be Riemannian under these conditions. However, in the context of non-Riemannian manifolds, as those with torsion present, a more general asymmetric connection $\Tilde{\Gamma}^{\kappa}_{\mu \nu}$ must be used. The torsion may then be defined as the asymmetry of the connection as the lower two indices are interchanged \cite{2013AnP...525..339M}
\begin{equation}
    T^{\kappa}_{\mu \nu} = \Tilde{\Gamma}^{\kappa}_{\mu \nu} - \Tilde{\Gamma}^{\kappa}_{\nu \mu}.
\end{equation}
This property has far reaching consequences for Maxwell's equations. To begin with, the field strength tensor $\Tilde{F}_{\mu \nu}$ has a more complex structure since it is no longer invariant under the change of metric as in Eq.~\eqref{invariant}. Instead the field strength in its covariant and contravariant basis, respectively, is given by
\begin{align}\label{Ftorsion}
    \Tilde{F}_{\mu \nu} = \nabla_{\mu} A_{\nu} - \nabla_{\nu} A_{\mu} =\nonumber \\ \partial_{\mu} A_{\nu} - \partial_{\nu} A_{\mu} + \left(\Tilde{\Gamma}^{\alpha}_{\mu \nu} - \Tilde{\Gamma}^{\alpha}_{\nu \mu}\right)A_{\alpha}
\end{align}
and 
\begin{equation}
    \Tilde{F}^{\mu \nu} = g^{\alpha \mu}g^{\beta \nu} \Tilde{F}_{\alpha \beta},
\end{equation}
and secondly, as already touched upon, the homogeneous equations are now sensitive to the background metric. In addition, due to the non-vanishing of the off-diagonal components of the metric, mixing of the field components in the contravariant basis is to be expected. Next we shall discuss the relationship between vorticity and torsion.

\subsubsection{Vorticity and torsion}

Let us consider the effective metric in the neighbourhood of a vortex \cite{1998JETPL..67..881V}. In this regime the condensate density is approximately constant but the superfluid flow $v_{\rm{sf}}$ is non-zero. Interestingly, this metric takes on an identical form to that of a spinning cosmic string \cite{Vickers_1987,Harari1988cb,PhysRevD.32.504,volovik1998vortex}. The vortex can thus be regarded as a torsion defect of the analogue space-time, which if encircled, leads to the accumulation of a geometric phase. This phase factor can be computed by interpreting the gravity as gauge theory in which the torsion plays the role of the field strength of the abelian tetrad fields $e^a_{\mu}$ which decomposes the metric as $g_{\mu \nu} = e^a_{\mu} e^b_{\nu} \eta_{a b}$, where $\eta_{a b}$ is the flat Minkowski metric and $e^a = e^a_{\mu} dx^{\mu}$. Setting the conformal factor $\Omega =1$, we may factorize the invariant line element as
\begin{equation}
    ds^2 = \left(c_{\rm{sf}}\gamma_{\rm{sf}}^{-1} dt + \frac{w \kappa}{c_{\rm{sf}}} \gamma_{\rm{sf}} d\theta\right)^2 - dr^2 + \gamma_{\rm{sf}}^{2} r^2 d\theta^2,
\end{equation}
where $w$ is the winding number of the vortex, $\kappa$ is the circulation quantum and
\begin{equation}
    \gamma_{\rm{sf}} = \frac{1}{\sqrt{1-\frac{v_{\rm{sf}}^2}{c_{\rm{sf}}^2}}}
\end{equation}
is the superfluid Lorentz factor. Writing the space-time line element in this way allows us to easily work out the tetrads
\begin{equation}
    e^0 =  c_{\rm{sf}} \gamma_{\rm{sf}}^{-1} dt + \frac{w \kappa}{c_{\rm{sf}}} \gamma_{\rm{sf}} d \theta, \ \ e^1 = dr \ \ \textrm{and} \ \ e^{2} = \gamma_{\rm{sf}} r d\theta.
    \label{exps}
\end{equation}
As already argued by Volovik \cite{volovik1998vortex}, a quasi-particle propagating past a vortex will experience a time delay due to the non-vanishing torsion within the vortex core. The torsion may then be probed by computing the holonomy induced as a quasi-particle encircles the vortex along a loop $\mathcal{C}$ \cite{Anandan1994TopologicalAG}. In its most generic form, the geometric phase acquired by the quasi-particle is given by $e^{i \phi}$ where
\begin{equation}
    \phi = \oint_{\mathcal{C}} \left(e_{\mu}^a P_a + \omega^b_{a \mu}M^a_b\right)dx^{\mu},
\end{equation}
$\omega^b_{a \mu}$ is the spin connection and $P_a$ and $M^a_b$ are the generators of translations and rotations, respectively. Now, since translations are commutative, we may factor out the component corresponding to time translation. The phase purely due to torsion can thus be obtained by integrating the $\theta$ 1-form in $e^0$ along $\mathcal{C}$, which gives us 
\begin{equation}
    \phi_t = \frac{w \kappa}{c_{\rm{sf}}} \oint_{\mathcal{C}} \gamma_{\rm{sf}} d\theta = \frac{2\pi w \kappa}{c_{\rm{sf}}} \gamma_{\rm{sf}}.
\end{equation}
Thus, we may interpret the Lorentz factor as the torsion density over a closed loop. Note also that the $2 \pi$ phase winding is recovered in the long distance limit as $\lim_{r\longrightarrow \infty} \gamma_{\rm{sf}} (r) = 1$. The relationship between torsion and the tetrad fields is captured by Cartan's first structure equation \cite{2018IJGMM..1540005O}
\begin{equation}
    T^a = de^a + \omega^a_b \wedge e^b.
\end{equation}
The solutions for the spin-connection components have already been worked out for the metric governing a spinning cosmic string \cite{Anandan1994TopologicalAG}, which in the case of a vortex thus are $\omega^1_{\theta 2}=-\omega^2_{\theta 1} = \gamma_{\rm{sf}}$, due the the equivalence between the two systems. The asymmetric connection coefficients, given a tetrad basis, can now be computed as \cite{2013AnP...525..339M}
\begin{equation}\label{torsioncon}
    \Tilde{\Gamma}^{\lambda}_{\mu \nu} = e^{\lambda}_a \partial_{\mu} e^a_{\nu} + e^{\lambda}_{a} e^{b}_{\nu} \omega^a_{\mu b}.
\end{equation}
Contrary to the Levi--Civita connection, this connection takes into account the effects on parallel transport due to torsion. In the following sections we shall use this object to couple the field strength to the vortex gravity.

\subsubsection{Gauss--Einstein law}
Going through the same procedure as in the curved case, Gauss's law becomes
\begin{align}
 \nabla_{\mu} \Tilde{F}^{\mu t} = \left(\partial_r + \Tilde{\Gamma}^r_{r r} + \Tilde{\Gamma}^{\theta}_{r \theta} + \Tilde{\Gamma}^{t}_{t r}\right) \Tilde{F}^{r t} + \nonumber \\
 \left(\partial_{\theta} + \Tilde{\Gamma}^r_{\theta r} + \Tilde{\Gamma}^{\theta}_{\theta \theta} + \Tilde{\Gamma}^{t}_{t \theta}\right) \Tilde{F}^{\theta t} + \\
 \left(\Tilde{\Gamma}^{t}_{r \theta} - \Tilde{\Gamma}^{t}_{\theta r}\right)\Tilde{F}^{r \theta}.
\end{align}
Already at this stage something interesting occurs: it seems like the electric field in Gauss's law defined on this metric is accompanied by a magnetic field $\Tilde{F}^{r \theta}$, due to the asymmetry of $\Tilde{\Gamma}^t_{r \theta}$. However, this should not be too surprising since the current in its contravariant basis has, due to the non-zero off-diagonal elements in the metric, mixed components 
\begin{equation}\label{torsioncurrent}
    J_{\rm{ME}}^{t} = g^{t \alpha} J^{\rm{ME}}_{\alpha} = \rho^{\rm{ME}} - \frac{v_{\rm{sf}}}{c_{\rm{sf}}}j^{\rm{ME}}_{\theta},
\end{equation}
given an azimuthal flow. This stems from the fact that when the source vortex is placed in a region in which the torsion is non-negligible, it will start to precess. In order to find the exact form of Gauss's law, we need to compute the asymmetric connection coefficients which yields
\begin{align}
    \nabla_{\mu} \Tilde{F}^{\mu t} = \left(\partial_r + \beta \partial_r \beta\right)\Tilde{F}^{r t} + \partial_{\theta}\Tilde{F}^{\theta t}- \nonumber \\ \left( (\partial_r - \gamma_{\rm{sf}})\alpha-\alpha\partial_r \beta \right)\Tilde{F}^{r \theta},
\end{align}
where $\alpha(r)=\frac{w \kappa}{c_{\rm{sf}}}\gamma_{\rm{sf}}(r)$ and $\beta(r) = r \gamma_{\rm{sf}}(r)$. Further, if we go to the rest frame of the vortex such that $v_{\rm{sf}}=0$, and consequently that $\alpha(r) = \frac{w \kappa}{c_{\rm{sf}}}$ and $\beta(r) = r$, the magnetic term vanishes so that Gauss's law, in the rest frame of the source, becomes (with unit coupling)
\begin{equation}
    \left(\partial_r + r\right)\Tilde{F}^{r t} + \partial_{\theta}\Tilde{F}^{\theta t} = \mu_0 \rho^{\rm{ME}},
\end{equation}
which is purely electric.

\subsubsection{Ampère--Einstein law}

Ampère--Einstein's law is obtained by letting $\nu=r,\theta$ in $\nabla_{\mu}\Tilde{F}^{\mu \nu}$, which yields the components
\begin{equation}
    \hat{\textbf{e}}_r: \     \ \nabla_{\mu}\Tilde{F}^{\mu r} = \partial_t \Tilde{F}^{t r} + \partial_{\theta} \Tilde{F}^{r \theta}
\end{equation}
and
\begin{equation}
    \hat{\textbf{e}}_{\theta}: \     \ \nabla_{\mu}\Tilde{F}^{\mu \theta} = \partial_t \Tilde{F}^{t \theta} + \partial_r \Tilde{F}^{r \theta}.
\end{equation}
Interestingly, at first sight, it appears that the torsion has no effect on this equation. This is not the case, however, since the contravariant field strength components are obtained by applying a non-diagonal metric, causing mixing to occur between the components, which thus yields 
\begin{equation}   
    \Tilde{F}^{t r}=\Tilde{F}_{t r} + \frac{v_{\theta}}{c_{\rm{sf}}}\Tilde{F}_{\theta r},\ \ \Tilde{F}^{t \theta}=\frac{v_{\rm{sf}}^2}{c_{\rm{sf}}^2}\left(\Tilde{F}_{\theta t} + \Tilde{F}_{r t}\right), \ \ \Tilde{F}^{r \theta}=\frac{v_{\rm{sf}}^2}{c_{\rm{sf}}^2}\Tilde{F}_{r \theta},
\end{equation}
 where $\Tilde{F}_{\mu \nu}$ is defined as in Eq.~\eqref{Ftorsion}.

\subsubsection{Einstein--Faraday and Einstein--Gauss \textbf{B}-law}

As already argued, the asymmetry of the torsional connection implies that the connection terms in the covariant derivative generally do not vanish, as is the case in the absence of torsion. Nonetheless, the terms containing the partial derivatives do vanish trivially in $(2+1)$-dimensions. The equation we are left with for the sourceless part thus is
\begin{equation}
    \left(\Tilde{\Gamma}^{\alpha}_{\mu \gamma} - \Tilde{\Gamma}^{\alpha}_{\gamma \mu}\right) \Tilde{F}_{\alpha \nu}\\ \nonumber + \left(\Tilde{\Gamma}^{\beta}_{\nu \gamma} - \Tilde{\Gamma}^{\beta}_{\gamma \nu}\right) \Tilde{F}_{\beta \mu} + \left(\Tilde{\Gamma}^{\delta}_{\nu \mu} - \Tilde{\Gamma}^{\delta}_{\mu \nu}\right) \Tilde{F}_{\delta \gamma}=0.
\end{equation}
This equation encompass information that is fundamentally different from that contained in the conventional homogeneous equations; it says nothing about the existence of magnetic monopoles, and nor does it assert anything pertaining to electromagnetic induction. Rather it stipulates how the field strength components are related to one another via the space-time geometry. We shall leave the derivation of the exact form of these equations for future work.
\section{Discussion}
A planar superfluid described by a scalar field has been studied. By generalizing the vortex electrodynamics picture put forth in \cite{2020PhRvA.101f3616S} to systems with non-constant densities, we have shown that the resulting Maxwell's equations can be viewed as a Maxwell theory minimally coupled to a gauge field $v_{\nabla |\Psi|}$ defined in Eq.~\eqref{v_d}. A gravity description was then adopted in which we considered the effective space-time defined by the analogue metric Eq.~\eqref{metric}. By treating this analogue space-time as the background on which Maxwell's equations live, we recovered the same set of Maxwell's equations as in the hydrodynamics description, given that an appropriate 3-current was introduced. As such, the core result of this work is the establishment of a correspondence between the hydrodynamics of a planar superfluid and Maxwell--Einstein theory, which is summarized in the following proposition:\\ \\
\textbf{Proposition} (Maxwell--Einstein-superfluidity correspondence)\\
\textit{Let $g_{\mu \nu}(\bf{r})$ be the metric of a $(d+1)$-dimensional conformal space-time and let $F_{\mu \nu}$ be the field strength tensor in $(2+1)$-dimensions, then the resulting Maxwell--Einstein theory is equivalent to the hydrodynamics of a planar superfluid, given that the following relationships are enforced}
\begin{align*}
    x &= \frac{2}{d-1},\\
    g_E = x+1 &= \frac{d+1}{d-1} \textrm{ and}\\
    g_M = x &= \frac{2}{d-1}   
\end{align*}
\textit{between the dimensionless parameter $x$, the dimension $d$ and the electromagnetic couplings $g_E$ and $g_M$.}\\ \\
For the sake of generality, we expanded upon this picture by considering the effective metric in the neighbourhood of a vortex. In this case, the torsion tensor had non-vanishing components which required us to introduce an asymmetric connection. Coupling to a torsional connection has, as underscored in this work, several interesting consequences for the resulting electrodynamics. Firstly, the field strength tensor takes on a fundamentally different form since the terms accounting for the non-trivial metric are non-vanishing, as is the case when only curvature is present. Secondly, the contravariant components of the field strength tensor are generally formed by superpositions of various components since the metric has off-diagonal elements. A third interesting consequence of the presence of torsion is that the homogeneous equations are no longer insensitive to the metric. If the connection only has curvature, the resulting homogeneous equations are left invariant, that is, they are still defined by the Bianchi identity. By contrast, these equations possess a fundamentally different structure in the torsional case. In particular, they are reduced from being of a differential-type to a simple linear-type set of equations, since all of the derivatives cancel out. As for future work, an interesting direction of study would be to consider condensates with spin. It has been shown that spinor condensates can be used to simulate e.g. quantum chromodynamics \cite{2000AcPPB..31.2837V,1997cond.mat.11031V,2001PhR...351..195V,2018PhRvA..97b3613E,tylutki2016confinement,cirac2010cold}. Coupling the Yang--Mills field strength tensor to the analogue connection may thus make possible the study of the effect of a non-vanishing curvature (and torsion) on analogue quark interactions.

\section*{Acknowledgements}
 I wish to thank Tapio Simula for many illuminating discussions on the topic of analogue physics in superfluids. I am also grateful to Mathis Moes for providing assistance in proof reading the manuscript. This research was supported by the Australian Research Council Future Fellowship FT180100020, and was funded by the Australian Government.

\bibliography{main.bbl}

\end{document}